\documentstyle[11pt]{article}
\begin{document}
\newcommand{\tr}{\bigtriangleup}
\newcommand{\be}{\begin{equation}}
\newcommand{\ee}{\end{equation}}
\newcommand{\ba}{\begin{eqnarray}}
\newcommand{\ea}{\end{eqnarray}}

\begin{center}
{\large \bf Asymptotic behavior in the scalar field theory}\\
\vspace*{5 true mm}
{\bf V.E. Rochev\footnote{E-mail address: rochev@ihep.ru}}\\
{\it Institute for High Energy Physics, 142280 Protvino, Moscow region, Russia}
\end{center}
{\small{ {\bf Abstract.} An asymptotic solution of the system of
Schwinger-Dyson equations for four-dimensional Euclidean scalar field theory
with interaction $\frac{\lambda}{2}(\phi^*\phi)^2$ is obtained.
For $\lambda>\lambda_{cr}=16\pi^2$  the two-particle amplitude has
the pathology-free asymptotic behavior at large momenta.
For $\lambda<\lambda_{cr}$ the amplitude  possesses Landau-type
singularity.}\\


\section{Introduction}
The problem of asymptotic behavior at  large momenta (or, at
short distances) is one of oldest problem of quantum field theory.
In general, this problem has been solved only for the
asymptotically-free models. A solution of the problem of
asymptotic behavior for other strictly renormalized models --
quantum electrodynamics, self-interacting scalar field, Yukawa
interaction -- requires to go out the framework of weak-coupling
approximation.

The first attempt to define the asymptotic behavior in quantum
electrodynamics was made by Landau and coworkers in the 1950s \cite{Landau}.
 The result was unrehearsed: the photon
propagator included the non-physical pole in Euclidean region of
momenta. Then similar poles were indicated in other strictly
renormalized models: self-interacting scalar field and Yukawa
interaction. The presence of such singularities in Euclidean
region  violates general principles of the local field theory and
is a serious problem for these models.

A method of removing these non-physical singularities  was proposed soon in works
\cite{Bog}. This method is based on the application  of
 Kallen--Lehmann representation to restore the correct analytical
structure and remains the basic tool to solve the problem
 of the non-physical singularities  to our time. However, the absence of
dynamical foundation   enforced to search a more detailed investigation
for the role of these singularities.

Further development of the quantum field theory  has demonstrated
that these non-physical singularities arise practically
 inevitably in the framework of any known non-perturbative methods:
at the renormalization-group summation,
  in the frameworks of  $1/N$-expansion and mean-field expansion, etc.
\footnote{See review \cite{Callaway} for the historical survey and further
references.}

A widespread opinion is formulated as a mathematical inconsistency
of the quantum field models that are not asymptotically free.
There is rigorous theorem \cite{Triviality} that the
four-dimensional scalar field theory with $\phi^4$ interaction on
the lattice does not  have an interacting continuum theory as its
limit for zero lattice spacing, i.e. the theory is trivial.
Weinberg argues \cite{Weinberg}, however, that this argument  is
inconclusive due to an  uncertainty   of the continuous limit in
this model.

Today the situation with triviality of  $\phi^4$ theory is vague
as before. Recent papers (see, e.g., \cite{Suslov},  \cite{Weisz},
\cite{Podolsky} ) in this topic maintain incompatible statements.
While  works \cite{Weisz} confirm the triviality scenario,
 works \cite{Suslov},
\cite{Podolsky} state non-trivial behavior in the strong-coupling
limit.

 The asymptotic short-distance region in the models without
asymptotic freedom is the region of strong (or, more exactly,
non-weak) coupling, and it is the main difficulty in its
investigation. It seems that the above-mentioned standard methods
are too tethered to the weak-coupling region and are not enough
meaningful to inform us about behavior at short distances  in
these models. In this paper the new approximation for the $\phi^4$
theory is investigated. This approximation is based on a system of
Schwinger-Dyson equations (SDEs) for the propagator and the two-particle function.
The first equation is the exact SDE for the propagator,
 and the second equation
is a truncated SDE for the two-particle function
("two-particle approximation"). This approximation is considered
as a first non-trivial step of the sequence of general
$n$-particle approximations, which tends to the exact infinite
system of SDEs at $n\rightarrow\infty$. An
investigation of some truncation of SDEs
 is not a news, of course. A new step is the consideration of a system of
these equations instead of the usual consideration of some single truncated equation.

A structure of the paper is as follows: in  section 2,
the necessary notations and definitions are given; SDE
 for the
generating functional of Schwinger functions is introduced in the
formalism of a bilocal  source. Using of the bilocal source
 is an essential point of the construction.
 We consider using this presumably as a convenient choice
of the functional variable. In particular, this variable is very convenient
for the construction of the mean-field expansion (MFE) by method of
work \cite{Ro1}, which is presented in section 2. The existence of the Landau
pole in the two-particle
amplitude of the leading approximation of MFE is also demonstrated in this section.

 In  section 3, a general
construction of the approximation scheme for the system of SDEs is given.
 In  section 4,  an alternative view to
the approximation of preceding section is given. The system of
equations for the propagator and the two-particle function is
considered as a basic approximation for the construction of the
modification of MFE of section 2. In section 5, the renormalization
of the system of equations is made and some supplement
simplifications are discussed.

 In section 6, the asymptotic solution of the
equation for the two-particle amplitude  at large momenta  is
presented.
 In section 7, the asymptotic behavior of the amplitude  at large momenta
is discussed. The amplitude in this model possesses a
self-consistent behavior (as a constant plus a decreasing
oscillating term) at the values of the  renormalized coupling
$\lambda\in (\lambda_{cr}, 2\lambda_{cr})$, where
$\lambda_{cr}=16\pi^2.$
 At $\lambda<\lambda_{cr}$ the amplitude has some Landau-type singularity
in the pre-asymptotic region. At $\lambda>2\lambda_{cr}$ the
method of solution is, strictly speaking, cannot be applied due to
the negative value of the field-renormalization constant, but if
one assumes formally such values, the results of
the permitted interval can be continued to the region of strong
coupling.
  Conclusions are presented in section 8.

\section{Preliminaries. SDE for the
generating functional, mean-field expansion and Landau pole}

Consider the  theory of a complex scalar field $\phi(x)$ in a
four-dimensional Euclidean space $(x \in E_4)$ with the Lagrangian
\be {\cal L} = -\partial_\mu
\phi^*\partial_\mu\phi-m^2_0\phi^*\phi-\frac{\lambda}{2}
(\phi^*\phi)^2
 \label{lagrangian} \ee
in the symmetric phase ($m_0^2>0, \, \lambda>0$).
 The generating functional of
$ 2n$-point ($n$-particle)
 Schwinger functions  can be written as a functional integral
\be G = \int D(\phi,\phi^*) \exp\bigg\{\int dx {\cal L}(x)-\int
dxdy \phi^*(x)\eta(x,y)\phi(y)\bigg\}, \label{G} \ee where
$\eta(x,y)$ is a bilocal source. The $n$th derivative of $G$ over
$\eta$ with the source being switched off is the $n$-particle
Schwinger function. The propagator of the field $\phi$ is $$
\Delta(x-y)=<\phi(x)\phi^*(y)>=-\frac{\delta
G}{\delta\eta(y,x)}\bigg\vert_{\eta=0}, \label{delta} $$
 the two-particle function is the second derivative of $G$,
etc.\footnote{A formalism  of the bilocal source in the quantum field theory was first
elaborated by Dahmen and Jona-Lasinio \cite{Dahmen}. 
For further development and references see, e.g.,  \cite{Mir}}

 The translational invariance of the functional integration measure leads to relation
$$
\int D(\phi,\phi^*) \frac{\delta}{\delta\phi^*(x)}\phi^*(y)
\exp\bigg\{\int dx {\cal L}(x)-\int dxdy
\phi^*(x)\eta(x,y)\phi(y)\bigg\}=0,
$$
which can be rewritten as the functional-differential
SDE for generating functional $G$: \be
(m_0^2-\partial^2_x)\frac{\delta G}{\delta\eta(y,x)}+\int
dy_1\eta(x,y_1) \frac{\delta G}{\delta\eta(y,y_1)}+\delta(x-y)G=
\lambda\frac{\delta^2G}{\delta\eta(x,x)\delta\eta(y,x)}.\label{SDE}
\ee To construct the mean-field expansion for the generating
functional we consider as a leading approximation the following
equation \be
\lambda\frac{\delta^2G^{(0)}}{\delta\eta(x,x)\delta\eta(y,x)}-
(m_0^2-\partial^2_x)\frac{\delta
G^{(0)}}{\delta\eta(y,x)}-\delta(x-y)G^{(0)}=0, \label{LOMFE} \ee
whose solution is (in operator notations)
$$
G^{(0)}=\exp\{- \Delta_0\cdot\eta\},
$$
where $\Delta_0(p)=(m^2+p^2)^{-1}$ and $ m^2=m_0^2+\lambda
\Delta_0(x=0).$ Here $\Delta_0(x=0)\equiv\int_\Lambda
\frac{d^4p}{(2\pi)^4}\Delta_0(p) $, and label $\Lambda$ means that
some regularization implied.

The leading approximation generates linear iteration scheme for SDE (\ref{SDE})
$$G=G^{(0)}+G^{(1)}+\cdots+G^{(n)}+\cdots,$$ where
\be
\lambda\frac{\delta^2G^{(n)}}{\delta\eta(x,x)\delta\eta(y,x)}-
(m_0^2-\partial^2_x)\frac{\delta G^{(n)}}{\delta\eta(y,x)}-\delta(x-y)G^{(n)}=\int dy_1
\eta(x,y_1)\frac{\delta G^{(n-1)}}{\delta\eta(y,y_1)}. \label{Iteration}
\ee

An idea of this  iterative scheme is as follows:
 we shall consider "equation with constant coefficients" (\ref{LOMFE})
as a leading approximation,
i.e.,  equation (\ref{SDE}) with the second term omitted. This term contains
the source $\eta$ manifestly. The Schwinger functions are the derivatives
of $G(\eta)$ in zero and only the behavior of $G$ near $\eta = 0$ is essential,
therefore such an approximation seems to be acceptable. The term omitted
 contains the source and should be treated as a
perturbation.

The general solution of equation (\ref{Iteration}) is the
functional $G^{(n)}= P_{2n}G^{(0)},$ where  $P_{2n}$ is a
polynomial of $2n$th degree on source $\eta$. Therefore at the
$n$th step the computation of Schwinger functions reduces to
 solving  a closed system of $2n$ linear integral equations.\\
The first-step  generating functional $G^{(1)}$ reads
$$
G^{(1)}=\bigg(\frac{1}{2}G_2\cdot\eta^2-\Delta_1\cdot\eta\bigg)G^{(0)}.
$$
Here $G_2$ is the leading-order two-particle function and $\Delta_1$ is the
next-to-the-leading-order correction to the propagator.

The iteration equation at $n=1$ gives the equation for the two-particle function
\be
(m^2-\partial^2_x)G_2\left(
\begin{array}{cc} x&y\\x'&y'\end{array}
\right)=\delta(x-y') \Delta_0(x'-y)-\lambda \Delta_0(x-y)G_2\left(
\begin{array}{cc} x&x\\x'&y'\end{array}
\right), \label{G2L} \ee whose solution is
 \ba G_2\left(
\begin{array}{cc} x&y\\x'&y'\end{array}
\right) = \Delta_0(x-y')\Delta_0(x'-y)- \nonumber \\ -\int dx_1
dx_2
\Delta_0(x-x_1)\Delta_0(x'-x_2)f(x_1-x_2)\Delta_0(x_1-y)\Delta_0(x_2-y'),
\label{G2MFE} \ea where in the momentum space \be
f(p)=\frac{\lambda}{1+\lambda L_0(p)}, \label{fMFE} \ee and \be
 L_0(p)=\int_\Lambda \frac{d^4q}{(2\pi)^4}\,
\Delta_0(p+q)\Delta_0(q) \label{LMFE}
\ee
is the single scalar loop.

Note, that first  term in formula (\ref{G2MFE})  is the missed
disconnected part of the two-particle function of a leading approximation.
Hence, the connected structure of the two-particle function is restored
 at the first step of iterations. Such a peculiarity of the
iteration scheme is originated by the bilocal source and is not
something exceptional: as is well-known, the similar phenomenon
appears also in constructing $1/N$-expansion in the bilocal source
formalism. The crossing properties of two-particle function in
such iteration schemes are also restored stage-by-stage when the
next terms in the expansion are taken into account (a
discussion of this topic see in work \cite{Ro2}).

All above formulae for the propagator $\Delta_0$ and the amplitude $f$ contain
divergent integrals and should be renormalized. In correspondence with the standard
recipe we introduce the renormalized Lagrangian
\be
{\cal L} = -\partial_\mu \phi^*\partial_\mu\phi-m^2\phi^*\phi-\frac{\lambda}{2}
(\phi^*\phi)^2, \label{lagrR}
\ee
where $\phi, m$ and $\lambda$ are the renormalized field, mass and coupling,
and add the counter-terms
\be
\Delta{\cal L} = -(z-1)\partial_\mu \phi^*\partial_\mu\phi
-\delta m^2\phi^*\phi-(z_\lambda-1)\frac{\lambda}{2}
(\phi^*\phi)^2, \label{counterterms}
\ee
which absorb the divergences.

Full  Lagrangian ${\cal L}_b={\cal L}+\Delta{\cal L}$ can be written as
\be
{\cal L}_b=-\partial_\mu \phi^*_b\partial_\mu\phi_b
-m^2_b\phi^*_b\phi_b-\frac{\lambda_b}{2}
(\phi^*_b\phi_b)^2, \label{lagrB}
\ee
where
\be
\phi_b=\sqrt{z}\phi, \;\;   m^2_b=\frac{m^2+\delta m^2}{z};\;
 \lambda_b=\frac{\lambda z_\lambda}{z^2}.\label{bare}
\ee
Then all above calculations are reproduced with  Lagrangian (\ref{lagrB}), i.e.,
with the replacement $m^2_0\rightarrow m^2_b, \; \lambda\rightarrow\lambda_b, \;
\Delta_0\rightarrow\Delta_b, \; G_2\rightarrow G_{2b}, \; f\rightarrow f_b$ etc.,
and the normalization conditions are imposed on the renormalized propagator $\Delta$
and amplitude $f$.
For the easement of the following calculations we choose the normalization point at zero
momenta.

The normalization conditions for the propagator $\Delta(p^2)=z^{-1}\Delta_b(p^2)$ are
\be
\Delta^{-1}(0)=m^2 \label{normProp1},
\ee
\be
\frac{d}{dp^2}\Delta^{-1}|_{p^2=0}=1 \label{normProp2}.
\ee
These conditions define the mass-renormalization counter-term $\delta m^2$ and the field-renormalization
constant $z$. It is easy to see that $z=1$ in the case, and, consequently,
$ \Delta^{-1}=\Delta^{-1}_b=m^2+p^2, \; G_2=G_{2b}$. Thus,  the amplitude
$f=f_b$ is
\be
f(p)=\frac{\lambda z_\lambda}{1+ \lambda z_\lambda L_0(p)}.
\ee
The normalization condition for the amplitude is
\be
f(0)=\lambda. \label{normAmp}
\ee
This condition defines the coupling-renormalization constant $z_\lambda$ and the renormalized
amplitude
\be
f (p)=\frac{\lambda}{1+\lambda L_r(p)}, \label{fRMFE}
\ee
where
\be
 L_r(p)=L_0(p)-L_0(0)=
-\frac{1}{16\pi^2}\int_0^1 dz \log(1+z(1-z)\frac{p^2}{m^2}) \label{Rloop}
\ee
is the renormalized loop.

As it follows from equations (\ref{fRMFE}) and (\ref{Rloop}) the
renormalized amplitude $f$ possesses a non-physical singularity
(Landau pole) in the point $p^2=M_L^2$, where $M^2_L$ is a
solution of the equation
$$
1+\lambda L_r(M^2_L)=0. \label{Landaupole}
$$
This equation has a solution at any positive $\lambda$
  since function $L_r$ covers all values from $0$ to $-\infty$. At $p^2\rightarrow\infty$
$$
L_r\simeq -\frac{1}{16\pi^2}\log\frac{p^2}{m^2}
$$
and $M_L\cong m \exp\{ \frac{8\pi^2}{\lambda}\}.$ As  was yet
noted in the introduction, the  same Landau pole arises at the
calculations of renormalized amplitude by other methods: in the
frameworks of  $1/N$-expansion and   renormalization-group
summation.

\section{The system of SDEs and two-particle approximation}

From SDE (\ref{SDE}) one can go to the
functional $Z=\log G$. The SDE for $Z$ is
\be
 (m_0^2-\partial^2_x)\frac{\delta Z}{\delta\eta(y,x)}+\int dy_1\eta(x,y_1) \frac{\delta
Z}{\delta\eta(y,y_1)}+\delta(x-y)=\lambda\bigg[
\frac{\delta^2Z}{\delta\eta(x,x)\delta\eta(y,x)}+ \frac{\delta
Z}{\delta\eta(x, x)}\, \frac{\delta Z}{\delta\eta(y,x)}\bigg]
\label{SDEZ} \ee The system of SDEs is an infinite set
of equations for $n$-particle functions $Z_n\equiv \delta^n
Z/\delta\eta^n|_{\eta=0}$. The first SDE is simply
equation (\ref{SDEZ}) with the source being switched off: \be
(m^2-\partial^2_x) \Delta(x-y)=\delta(x-y)-\lambda Z_2\left(
\begin{array}{cc} x&y\\x&x\end{array}
\right),  \label{EqDelta}
\ee
where $m^2=m_0^2+\lambda\Delta(x=0)$ and
 $\Delta\equiv -\delta Z/\delta\eta|_{\eta=0}$ is the propagator (or, the
one-particle function). The second SDE is the derivative of
(\ref{SDEZ}) with the source being switched off:
\ba
(m^2-\partial^2_x)Z_2\left(
\begin{array}{cc} x&y\\x'&y'\end{array}
\right)=\delta(x-y') \Delta(x'-y)-
 \nonumber \\
-\lambda \Delta(x-y)Z_2\left(
\begin{array}{cc} x&x\\x'&y'\end{array}
\right) +\lambda Z_3\left(
\begin{array}{ccc}x&y\\x&x\\x'&y'\end{array}
\right), \label{Dyson2} \ea and so on. The $n$th SDE
is the $(n-1)$th derivative of SDE (\ref{SDEZ}) with the source
being switched off and includes a set of functions from
one-particle function $\Delta$ to $(n+1)$-particle function
$Z_{n+1}$.
 We call "the $n$-particle approximation of the system of SDEs"
the system of  $n$ SDEs in which the first $n-1$ equations are exact and $n$th
SDE is truncated by omitting the $(n+1)$-particle function. It is evident that
the sequence of such approximations goes to the exact set of SDEs at $n\rightarrow\infty$.
The one-particle approximation is simply equation (\ref{EqDelta}) without $Z_2$. This approximation
has a trivial solution which is a free propagator. The two-particle approximation is a system of equation
(\ref{EqDelta}) and equation (\ref{Dyson2}) without $Z_3$:
\be
(m^2-\partial^2_x)Z_2\left(
\begin{array}{cc} x&y\\x'&y'\end{array}
\right)=\delta(x-y') \Delta(x'-y)-\lambda \Delta(x-y)Z_2\left(
\begin{array}{cc} x&x\\x'&y'\end{array}
\right),  \label{EqG2}
\ee
which includes $\Delta$ and two-particle function $Z_2$. This non-linear system will be
the  object of present investigation.

The idea of the present approximation scheme is very simple and
natural. However, the calculations became more and more
complicated at each following stage: e.g., the three-particle
approximation is a system of three non-linear equations for the
propagator, the two-particle function and the three-particle
function. In following section  some other scheme will be
considered, which leads to same equations (\ref{EqDelta}) and
(\ref{EqG2}), but is more realistic in the calculational sense.

\section{The
particular solution of SDE for the
generating functional and modified
mean-field expansion}

Another view to the origin of system (\ref{EqDelta}) and
(\ref{EqG2}) is based on a modification of  the mean-field
expansion of  section 2
 with taking into account a particular
solution of functional-derivative equation (\ref{SDEZ}). It is
easy to see that SDE (\ref{SDEZ})   has the simple particular
solution \be
Z_{part}(\eta)=\frac{1}{2}Z_2\cdot\eta^2-\Delta\cdot\eta,
\label{Zpart} \ee where functions $Z_2\left(\begin{array}{cc}
x&y\\x'&y'\end{array} \right)$ and $\Delta(x-y)$ satisfy to the
system of equations (\ref{EqDelta}) and (\ref{EqG2}).

  This system should be supplemented by one more
equation for $Z_2$:
\ba
\lambda Z_2\left(
\begin{array}{cc} x&x\\x'&y'\end{array}
\right)Z_2\left(
\begin{array}{cc} x&y\\x''&y''\end{array}
\right)+\lambda Z_2\left(
\begin{array}{cc} x&x\\x''&y''\end{array}
\right)Z_2\left(
\begin{array}{cc} x&y\\x'&y'\end{array}
\right)= \nonumber
\\
=\delta(x-y')Z_2\left(
\begin{array}{cc} x'&y\\x''&y''\end{array}
\right)+\delta(x-y'')Z_2\left(
\begin{array}{cc} x''&y\\x'&y'\end{array}
\right). \label{G2G2} \ea The system of three equations
(\ref{EqDelta}), (\ref{EqG2}) and (\ref{G2G2}) for two functions
$\Delta$ and $Z_2$ are overfull and seemingly has not  physically
meaningful solutions. However, as it will be show below, equation
(\ref{G2G2}) does not play a role for the construction of the modified
mean-field expansion. Really, for this construction we define the
functional $\bar{G}$ as \be Z = Z_{part}(\eta)+\log\bar{G}. \ee
 Taking into account equations (\ref{EqDelta}) and (\ref{EqG2}) (but not (\ref{G2G2})!), we obtain
the SDE for the functional $\bar{G}$: \ba (m^2-\partial^2)\frac{\delta
\bar{G}}{\delta\eta}-\lambda\frac{\delta^2\bar{G}}{\delta\eta\delta\eta}
+\lambda\frac{\delta \bar{G}}{\delta\eta}\, \Delta+
\eta\cdot\frac{\delta \bar{G}}{\delta\eta}- \lambda\frac{\delta
\bar{G}}{\delta\eta} \,Z_2\cdot\eta-\lambda Z_2\cdot \eta
\,\frac{\delta \bar{G}}{\delta\eta}=
\nonumber \\
= \bar{G}\,\biggl[\lambda  Z_2\cdot \eta \,Z_2\cdot \eta-
\eta\cdot Z_2\cdot\eta\biggr] \label{Gbar} \ea
 (in operator
notation; the central dot is an integration). If we require also to
satisfy equation (\ref{G2G2}), then rhs of equation
(\ref{Gbar}) is zero, and we obtain the situation, which in the
theory of ordinary differential equations is named as "the
reduction of order": instead of the second-order linear equation
(\ref{SDE}), we have a first-order linear equation for
$\delta\bar{G}/\delta\eta$. But the construction of the mean-field
expansion for $\bar{G}$, which is similar to the mean-field
expansion for $G$ of section 2, absolutely has not need of such
reduction. For this reason we shall not require a satisfaction of
equation (\ref{G2G2}).

To construct the mean-field expansion for $\bar{G}$ we consider as a leading-order
 approximation (in full correspondence with the principle of the construction of
 mean-field expansion
of section 2)   the terms of equation (\ref{Gbar}), which do not
contain the source $\eta$ manifestly, i.e. the leading-order
equation is \be (m^2-\partial^2_x)\frac{\delta
\bar{G}_0}{\delta\eta(y,x)}-
\lambda\frac{\delta^2\bar{G}_0}{\delta\eta(x,x)\delta\eta(y,x)}+\lambda\frac{\delta
\bar{G}_0}{\delta\eta(x,x)} \Delta(x-y)=0 \ee All other terms
(including the rhs of (\ref{Gbar})) are considered as the
perturbation.

This equation also has a linear exponent as a solution. Then, in
correspondence with section 2, we have the iteration scheme
$$
\bar{G}=\bar{G}^{(0)}+\bar{G}^{(1)}+\cdots+\bar{G}^{(n)}+\cdots,
$$
and the equation for $\bar{G}^{(n)}$ is
$$
(m^2-
\partial^2)\frac{\delta
\bar{G}_n}{\delta\eta}
-\lambda\frac{\delta^2\bar{G}_n}{\delta\eta\delta\eta}
+\lambda\frac{\delta \bar{G}_n}{\delta\eta}\,\Delta
 +
\biggl\{\eta\cdot\frac{\delta \bar{G}_{n-1}}{\delta\eta}
-\lambda\frac{\delta \bar{G}_{n-1}}{\delta\eta} \,Z_2\cdot
\eta-\lambda Z_2\cdot \eta\,\frac{\delta
\bar{G}_{n-1}}{\delta\eta}\biggr\}=
$$
$$
=\biggl[Z_2\cdot \eta\,Z_2\cdot \eta-\eta\cdot Z_2\cdot \eta
\biggr]\,      \bar{G}_{n-2}
$$
The general solution of this equation is the functional
$\bar{G}^{(n)}=\bar{P}^{(n)}\bar{G}^{(0)}$, where $\bar{P}^{(n)}$
is a polynomial on $\eta$. At $n$th step of this iteration scheme,
we have a closed system of linear integral equations, and
therefore this scheme is much less complicated in comparison to
the scheme of preceding section. Equations (\ref{EqDelta}) and
(\ref{EqG2}) are the basic approximation for this expansion.

\section{The renormalized equations of the two-particle approximation}

Equations (\ref{EqDelta}) and (\ref{EqG2}) are the system of non-linear equations
for the functions $\Delta$ and $Z_2$. In equation (\ref{EqG2}),
the  two-particle function $Z_2$ can be considered as a functional of $\Delta$, and the "solution"
of this equation can be easily found:
\ba
Z_2\left(
\begin{array}{cc} x&y\\x'&y'\end{array}
\right) =
\Delta_c(x-y')\Delta(x'-y)-
 \nonumber \\
-\int dx_1dx_2\Delta_c(x-x_1)\Delta(x'-x_2)f(x_1-x_2)\Delta(x_1-y)\Delta_c(x_2-y').
\label{SolG2}
\ea
Here,   function $f(p)$ in the momentum space is a solution of the equation
\be
\frac{1}{f(p)}=\frac{1}{\lambda}+\int \frac{ d^4q}{(2\pi)^4}
\Delta_c(p+q)\Delta(q),   \label{Eqf}
\ee
where
\be
\Delta_c(p)=\frac{1}{m^2+p^2}. \label{Deltac}
\ee
Taking into account equations (\ref{SolG2}) and (\ref{Eqf}), we obtain for $\Delta$
the following equation in the momentum space
\be
(m^2+p^2)\Delta(p)=1-\Delta(p)\int \frac{ d^4q}{(2\pi)^4} \Delta_c(p-q) f(q).
\label{DeltaNR}
\ee

Note that  a very crude approximation for equation (\ref{DeltaNR}) is
\be
(m^2+p^2)\Delta(p)\approx 1 \;\Longrightarrow \; \Delta(p)\approx\Delta_c(p).
\ee
Within this approximation we obtain as a solution of  equation (\ref{Eqf})
the mean-field amplitude of equation (\ref{fMFE}). So the mean-field approximation and the
equivalent leading-order $1/N$--expansion are contained in the
two-particle approximation as an approximate solution.

The renormalization of  equations (\ref{Eqf}) and (\ref{DeltaNR})
 can be performed in correspondence with the
general recipe of section 2 by introducing  counter-term Lagrangian (\ref{counterterms})
and full Lagrangian (\ref{lagrB}). In terms of the bare quantities the bare propagator is
\be
\Delta^{-1}_b(p^2)=m_b^2+\lambda_b\Delta_b(x=0)+p^2+\Sigma_b(p^2),
\label{DeltaB}
\ee
where
\be
\Sigma_b(p^2)=\int \frac{d^4q}{(2\pi)^4}\Delta_c(p-q) f_b(q)
\label{SigmaB}
\ee
is the bare mass operator.

The normalization conditions (\ref{normProp1}) and (\ref{normProp2}) for the
renormalized propagator $\Delta=z^{-1}\Delta_b$ define the mass-renormalization counter-term
$\delta m^2$ and the field-renormalization constant
\be
z=(1+ \Sigma'_b(0))^{-1}. \label{z}
\ee
 Then the equation for the renormalized propagator is
\be
(m^2+p^2)\Delta(p^2)=1-\Delta(p^2)\Sigma_r(p^2), \label{DeltaR}
\ee
where
\be
\Sigma_r(p^2)=z[\Sigma_b(p^2)-
\Sigma_b(0)-p^2\Sigma'_b(0)] \label{SigmaR}
\ee
is the renormalized mass operator.

The equation for $f_b$ is
\be
 \frac{1}{f_b(p)}=\frac{1}{\lambda_b}+L_b(p), \label{EqfB}
\ee
where
\be
L_b(p)=\int \frac{d^4q}{(2\pi)^4} \Delta_c(p+q)\Delta_b(q)=z\int
\frac{d^4q}{(2\pi)^4} \Delta_c(p+q)\Delta(q)  \label{loopB}
\ee
is the bare loop operator.

The renormalized amplitude $F$ is defined as an amputation of the connected
part $Z_2^{con}$ of the renormalised two-particle function $Z_2$:
$$
Z_2^{con} =
-\Delta\cdot\Delta\cdot F\cdot \Delta\cdot\Delta.
$$
In correspondence with the bare version of equation (\ref{SolG2})
$$
Z_2^{con}=z^{-2}Z_{2b}^{con}= -z^{-2}\Delta_c\cdot\Delta_b\cdot
f_b\cdot\Delta_b\cdot\Delta_c=-\Delta_c\cdot\Delta\cdot
f_b\cdot\Delta\cdot\Delta_c
$$
and we have (in momentum space):
$$
F\left(
\begin{array}{cc} p_x&p_y\\p'_x&p'_y\end{array}
\right)=\Delta^{-1}(p_x)\Delta_c(p_x) f_b(p_x+p_y) \Delta_c(p'_y)\Delta^{-1}(p'_y).
$$
The renormalized coupling $\lambda$ is defined as
$$
F\big|_{p_i=0} =\lambda,
$$
and due to the normalization condition (\ref{normProp1}) for the propagator we have
\be
f_b(0)=\lambda. \label{normfB}
\ee
This normalization condition  together with equation (\ref{EqfB}) defines the
 coupling-renormalization constant $z_\lambda$ and the renormalized equation for $f\equiv f_b$:
\be
\frac{1}{f(p^2)}=\frac{1}{\lambda}+L_r(p^2), \label{EqfR}
\ee
where
\be
L_r(p^2)=L_b(p^2)-L_b(0) \label{loopR}
\ee
is the renormalized loop operator.

Equations (\ref{DeltaR}) and (\ref{EqfR}) are the system of nonlinear integral
equations for the propagator and the amplitude. This system with taking into account the normalization
conditions (\ref{normfB}) and (\ref{normProp1})--(\ref{normProp2}) can be solved by the
expansion in the  vicinity of the point $p=0$. Such solution, however, is not interesting because
it is  some part of the usual perturbation theory over the coupling $\lambda$.

Much more interesting problem is to look for the asymptotic behavior of the solution at large
momenta.
In the large-momenta region, an essential technical  simplification is possible, namely, one can
replace in integrals (\ref{SigmaB}) and (\ref{loopB}) the function $\Delta_c$
(see (\ref{Deltac})) by massless function $1/p^2$:
\be
\int \frac{d^4q}{(2\pi)^4} \Phi(q^2)\,\Delta_c(p-q)\Longrightarrow\int \frac{d^4q}{(2\pi)^4}
\,\frac{ \Phi(q^2)}{(p-q)^2}.
\label{zeromass}
\ee
Then it is possible to use the well-known formula
\be
\int \frac{d^4q}{(2\pi)^4}
\,\frac{ \Phi(q^2)}{(p-q)^2}=\frac{1}{16\pi^2}\bigg[\frac{1}{p^2}
\int_0^{p^2} \Phi(q^2)\, q^2\, dq^2+\int_{p^2}^\infty \Phi(q^2)\,dq^2\bigg].
\label{Int}
\ee
The approximation (\ref{zeromass}) is quite usual in investigations in the deep-Euclidean region,
and formula (\ref{Int}) highly enables the calculations and, as a major point, permits us to go
from integral equations to differential ones (see below).

As an example of using  formula (\ref{Int}), we calculate the field-renormalization
constant $z$.
Due to the normalization condition (\ref{normfB}) we have $f\rightarrow\lambda$ at
$p\rightarrow 0$. Consequently, with the application of formula (\ref{Int}) we obtain
\be
\Sigma'_b(0)=-\frac{1}{2}\cdot\frac{\lambda}{16\pi^2},
\label{Sigma'}
\ee
and, taking into account (\ref{z}),
\be
z=\frac{2}{2-\lambda/16\pi^2}.
\ee
Due to the positivity of $z$, this formula  implies the limitation
$\lambda<32\pi^2\approx 320$ on all following calculations.

Further essential simplification of the system of equations for
$f$ and $\Delta$ consists in replacing of equation (\ref{DeltaR})
by the approximate relation \be (m^2+p^2)\Delta(p^2)\approx
1-\Delta_c(p^2)\Sigma_r(p^2). \label{appr} \ee 
This approximation
enables us "to unleash" the system, keeping at the time the
non-linearity, i.e., to obtain the non-linear integral equation
for $f$, which does not contain $\Delta$. This approximation, of
course, is less substantiated and should be considered as an
iteration of the initial equations. Some grounds for this
approximation can be obtained {\it a posteriori} if the
deep-Euclidean behavior of the propagator $\Delta$ does not
essentially differ  from the behavior of the free propagator.

Approximation (\ref{appr}) defines the equation for $f$, which will be a main object of the
following consideration. For the future convenience we introduce new quantities
$y, \;g$ and  dimensionless variable $t$ as
\be
y=\frac{z}{16\pi^2} \,f, \;\; g=\frac{z}{16\pi^2}\,\lambda, \;\; t=\frac{p^2}{m^2}.
\label{notation}
\ee
The normalization condition for $y(t)$ is
\be
y(0)=g.
\label{y(0)}
\ee

After a simple calculation, we obtain the equation for $y(t)$ in the
form \be \frac{1}{y}=\frac{1}{g}+l(t)+\int_0^t dt_1
K(t|t_1)\,y(t_1), \label{Eqy} \ee
 where \be
l(t)=\bigg(\frac{g}{2}-1\bigg)\log
(1+t)+(1-g)\big(1-\frac{1}{t}\log (1+t)\big), \label{l} \ee 
and
the kernel $K$ is \be
K(t|t_1)=\frac{t_1}{t}-1+\frac{1}{t}\log\frac{1+t}{1+t_1}
+t_1\log\frac{t(1+t_1)}{t_1(1+t)}. \label{K} \ee 
Integral equation
(\ref{Eqy}) is the non-linear Volterra equation and can be reduced
to the non-linear differential equation of the fourth order \be
\frac{d^2}{dt^2}\bigg[ t(t+1)^2\frac{d^2}{dt^2}\bigg(
\frac{t}{y}\bigg)\bigg]=g-2-y. \label{EqDiff} \ee 
It worth noting
that the differential equation (\ref{EqDiff}) will be used only for a
suggestion the form of the asymptotic solution of the integral
equation (\ref{Eqy}).

\section{Asymptotic solution for the amplitude}

As it easy to see, the differential equation (\ref{EqDiff}) at $g\neq
2$ has the exact solution \be y_{exact}=g-2. \label{exacty} \ee
This solution is not, however, the exact solution of the integral
equation (\ref{Eqy}), but the leading term of the asymptotic expansion
at $t\rightarrow\infty$. Indeed, a simple calculation demonstrates
that after the substitution of (\ref{exacty}) into (\ref{Eqy}) the
leading increasing logarithmic term is cancelled. Hence, the
asymptotic solution of the integral equation (\ref{Eqy}) at $g\neq 2$
has the form \be y=g-2 +\varphi(t), \label{yas} \ee 
where
$\varphi(t)\rightarrow 0$ at $t\rightarrow\infty$. The function
$\varphi(t)$ at large $t$ satisfies to the linear differential
equation \be \frac{d^2}{dt^2}\bigg[ t^3\frac{d^2}{dt^2}\bigg(
t\varphi\bigg)\bigg]=(g-2)^2\varphi. \label{Eqphi} \ee
 This
equation suggests the form of next-to-the-leading asymptotic term
$\varphi$. The solutions of this Euler-type equation are
$t^{\alpha-1}$, where \be \alpha^2(\alpha^2-1)=(g-2)^2.
\label{char} \ee 
This characteristic equation has two real and two
imaginary roots. The first real root is greater than 1, and
corresponds to  an increasing solution. The second real root is
less than --1, and corresponds to a fast-decreasing solution. The
imaginary roots correspond
to decreasing as $t^{-1}$ and oscillating solutions. Hence, just these roots define the
next-to-the-leading asymptotic term of the solution of integral equation (\ref{Eqy}).

For further calculation it is convenient to go to the variable \be
{\sc x}=\log(1+t) \label{x} \ee 
In terms of this variable the
next-to-the-leading asymptotic term  is \be \varphi=e^{-{\sc
x}}(A\cos \omega {\sc x}-B\sin\omega {\sc x}), \label{phias} \ee
where
 \be \omega= \sqrt{\sqrt{(g-2)^2+\frac{1}{4}}-\frac{1}{2}}.
\label{omega} \ee 
Substituting  (\ref{yas}) and
(\ref{phias}) into the integral equation (\ref{Eqy}) gives us at
$t\rightarrow\infty$ the relation \be
\frac{1}{g-2}=\frac{1}{g}-1-a(\omega)A+\omega b(\omega)B,
\label{AB} \ee where \be
a(\omega)=\sum_{k=1}^{\infty}\frac{1}{(k+1)(k^2+\omega^2)}, \;\;
b(\omega)=\sum_{k=1}^{\infty}\frac{1}{k(k+1)(k^2+\omega^2)}.
 \label{ab}
\ee
These sums can also be expressed in terms of the special functions (see, e.g. \cite{Prudnikov}).

Due to the normalization condition (\ref{y(0)}) we have $A=2$, and equation (\ref{AB})
defines the quantity $B(g)$.

\section{Asymptotic behavior of the amplitude}

At $t\rightarrow\infty$ we have
$
y\rightarrow g-2.
$
At $g<2$ such behavior contradicts to continuity.
If the function $y$ is continuous, then at $g<2 \;\; y(t_0)=0$ in some point  $t_0\in(0; +\infty)$
 (since $y$ changes the sign). But in the case the integral equation (\ref{Eqy}) is not fulfilled in
the point $t_0$.
Consequently the function $y$ is singular in some point
of pre-asymptotic region (Landau pole, or something similar), and at
  $g<2$ this model is inconsistent beyond the region of small momenta.

Hence, the region of applicability of this model is
 $g>2$, or
\be
\lambda_{cr}< \lambda <2\lambda_{cr},  \label{interval}
\ee
where
\be
\lambda_{cr}=16\pi^2\cong 160. \label{lambdacr}
\ee

The first two terms of asymptotic expansion in this interval of
$\lambda$ at large  $t=\frac{p^2}{m^2}$ are given as follows \be
f\simeq
2(\lambda-\lambda_{cr})+\frac{2\lambda_{cr}-\lambda}{t}\bigg(\cos(\omega\log
t)- \frac{B(\lambda)}{2}\sin(\omega\log t)\bigg), \label{fas} \ee
where
 \be
 \omega(\lambda)= \sqrt{\sqrt{\bigg(
\frac{4(\lambda-\lambda_{cr})}{2\lambda_{cr}-\lambda}\bigg)^2+\frac{1}{4}}-\frac{1}{2}},
\ee
 and \be
B(\lambda)=\frac{1}{\omega\,b(\omega)}\bigg[\frac{(2\lambda_{cr}-\lambda)^2}
{4\lambda(\lambda-\lambda_{cr})}+1+2a(\omega)\bigg], \ee 
At
$\lambda\rightarrow\lambda_{cr}$
 \be f\simeq
2(\lambda-\lambda_{cr})+\frac{C}{(\lambda-\lambda_{cr})^2}\,\frac{1}{t}\,
\sin(\omega\log t), \ee
 where $C=(2\pi)^6/b(0)$, i.e., in the
vicinity of the  critical point $\lambda_{cr}$ the amplitude of
oscillations tends to infinity and the solution is destroyed.

On the other end of the interval at $\lambda\rightarrow
2\lambda_{cr}$
 \be f\simeq
2\lambda_{cr}-4\pi\sqrt{2\lambda_{cr}-\lambda}\,\frac{1}{t}\,
\sin(\omega\log t), \ee 
i.e., \quad $f(t)\rightarrow f(0)$ -- the
amplitude at $t\rightarrow\infty$ tends to the initial value of
the point $t=0$. In other words, at $\lambda=2\lambda_{cr}$  the
asymptotic reconstruction of the amplitude occurred.

At $\lambda>2\lambda_{cr}$  the
method of solution is, strictly speaking, cannot be applied due to
the negative value of the field-renormalization constant $z$, but if
one formally assumes such values, the result of equation (\ref{fas})
 can be continued to the region of strong
coupling.

\section{Conclusions}
The most interesting feature of the result obtained is, of course,
the existence of the critical point $\lambda_{cr}$, which divides
the whole set of coupling values in two region: the weak-coupling
region of inconsistency and the strong-coupling region of
non-trivial self-consistent behavior.\footnote{Note that a singular behavior on
parameters for the 
theory of four-dimensional scalar field was marked before. Some years ago,
 using the local potential approach
 to the Wilson renormalization group,
Halpern and Huang \cite{Halpern}  have found a class of non-trivial interactions of a scalar field in
four dimensions. As pointed by Morris \cite{Morris}, such interactions correspond to
singular effective potential. The connection of this fact with our result is, however, unclear
due to very different methods of investigation.}  (The existence of the second
specific point  seems to be an artefact of the calculation
scheme.) This division of the coupling values cannot be indicated
by the expansions  tightly connected with perturbation theory
(e.g., the renormalization-group summation
  and the $1/N$-expansion). On the other hand, the strong-coupling
  limit is also not susceptible to the existence of such division,
  since in the framework of the Schwinger-Dyson formalism the
  strong-coupling expansion requires for the physically-acceptable
interpretation   some  subsidiary summation \cite{Bender, Ro3}
based on  unknown grounds.

\section*{Aknowlegements}
Author is grateful to the participants of IHEP Theory Division Seminar
for useful discussion.


\begin{thebibliography}{99}

\bibitem{Landau} Landau L D, Abrikosov A A and Khalatnikov I M 1954
{\it Dokl.Akad.Nauk Ser.Fiz.} {\bf 95}  1177;\\
Landau L D and Pomeranchuk I Ya 1955
{\it Dokl.Akad.Nauk Ser.Fiz.} {\bf 102}  489


\bibitem{Bog} Redmond P J 1958 {\it Phys.Rev.} {\bf 112} 1404;\\
 Bogoliubov N N, Logunov A A and Shirkov D V 1959
{\it ZhETF} {\bf 37} 805

\bibitem{Callaway} Callaway D 1988 {\it Phys.Reports} {\bf 167} 241

\bibitem{Triviality}
 Frohlich J 1982 {\it Nucl.Phys. B} {\bf 200} 281

\bibitem{Weinberg} Weinberg S  1995 {\it The Quantum Theory of Fields}
Vol. II, Ch. 18 (Cambridge Univ. Press)

\bibitem{Suslov}  Suslov I M 2008  {\it ZhETF} {\bf 134} 490 [{\it JETP} {\bf 107} 413];
   2010  {\it ZhETF} {\bf 138} 508 [{\it JETP} {\bf 111} 450]

\bibitem{Weisz}
Wolff  U 2009 {\it Phys.Rev. D} {\bf 79} 105002;\\
Weisz P and Wolff  U 2010
%{\it Triviality of $\phi^4_4$ theory: small volume expansion and new data}
 arXiv:1012.0404 [hep-lat]

\bibitem{Podolsky} Podolsky D I  2010
% {\it On triviality of $\lambda\phi^4$ quantum field theory in four dimensions}
 arXiv:1003.3670 [hep-th]

\bibitem{Dahmen}  Dahmen H D and Jona-Lasinio G 1967 {\it Nuovo Cim. A} {\bf52} 807

\bibitem{Mir}
 Miransky  V A  1993 {\it Dynamical Symmetry Breaking in
Quantum Field Theories} Ch. 8 (Singapore: World Scientific)

\bibitem{Ro1}
Rochev V E   1997 {\it J.Phys. A: Math. Gen.} {\bf 30} 3671

\bibitem{Ro2}  Rochev V E 2009 {\it Teor.Mat.Fiz.} {\bf 159} 81
 [{\it Theor.Math.Phys.} {\bf 159} 488]

\bibitem{Prudnikov} Prudnikov A P et al  
1986 {\it Integrals and Series}  Vol. 1
 (Gordon \& Breach Sci. Publ., New York)

\bibitem{Bender} Bender C M, Cooper F and Simmons L M 1989 {\it
Phys.Rev. D} {\bf 39} 2343

\bibitem{Ro3}
Rochev V E   1993 {\it J.Phys. A: Math. Gen.} {\bf 26} 1235

\bibitem{Halpern}
Halpern K and Huang K 1995 {\it Phys.Rev.Lett.} {\bf 74} 3526;
1996 {\it Phys.Rev. D} {\bf 53} 3252; {\it Phys.Rev.Lett.} {\bf 77} 1659

\bibitem{Morris} Morris T R 
  1996 {\it Phys.Rev.Lett.} {\bf 77} 1658



\end{thebibliography}
\end{document}